\documentclass[submission,copyright,creativecommons]{eptcs}
\providecommand{\event}{LAFM 2013} 
\usepackage{breakurl}             
\usepackage{graphicx}  		  

\title{BEval: A Plug-in to Extend Atelier B with Current Verification Technologies}
\author{Val\'{e}rio Medeiros Jr. \thanks{This work has been partially supported by CNPq grants 560014/2010-4, 308008/2012-0, 573964/2008-4 (National Institute of Science and Technology for Software Engineer - INES)}
\institute{IFRN\\ Natal, Brazil}
\institute{Federal Institute of Education, Science and Technology\\
of Rio Grande do Norte\\
Natal, Brazil}
\email{valerio.medeiros@ifrn.edu.br}
\and
David D\'{e}harbe \footnotemark[1]
\institute{UFRN\\ Natal, Brazil}
\institute{Federal University of Rio Grande do Norte\\
Department of Informatics and Applied Mathematics 
\\
Natal, Brazil}
\email{\quad david@dimap.ufrn.br}
}
\def\titlerunning{BEval: A Plug-in to Extend Atelier B with Current Verification Technologies}
\def\authorrunning{V. Medeiros Jr. \& D. D\'{e}harbe}
\begin{document}
\maketitle

\begin{abstract}
  This paper presents BEval, an extension of Atelier B to improve automation in
  the verification activities in the B method or Event-B. It combines a tool for
  managing and verifying software projects (Atelier B) and a model
  checker/animator (ProB) so that the verification conditions generated in the
  former are evaluated with the latter. In our experiments, the two main
  verification strategies (manual and automatic) showed significant improvement
  as ProB's evaluator proves complementary to Atelier B built-in provers. We
  conducted experiments with the B model of a micro-controller instruction set;
  several verification conditions, that we were not able to discharge
  automatically or manually with Atelier B's provers, were automatically
  verified using BEval.
\end{abstract}

\section{Introduction}

Classical B and Event-B are formal methods initially developed by
J.-R. Abrial~\cite{DBLP:books/daglib/0015096,AbrialBHHMV10} that contain the
notion of abstract machine and refinement. These methods are widely applied in
safety critical systems and supported by Atelier B~\cite{AtelierB} and others
tools. ProB~\cite{DBLP:dblp_conf/fm/LeuschelB03} is a tool for animation, model
checking as well as an expression evaluator.  A Rodin
plug-in~\cite{AbrialBHHMV10,LiBeLe07_219} to interact with ProB has been
developed and is used as a disprover.
The goal of the BEval project was to develop a similar plug-in for Atelier
B. This goal was driven by our attempt to streamline the verification of proof
obligations generated in the development of a formal model of a micro-controller
instruction set~\cite{Valerio_SBMF2012}. Indeed their verification with the
automatic prover available in Atelier-B was often inconclusive and required
time-consuming use of the interactive prover.

Currently, several components of Atelier B are neither open source nor free,
most notably the mathematical rule validator tool and the theorem prover for the
B method and Event-B are closed. Moreover the main Atelier B theorem prover
(krt) did not evolve significantly in the past decades. Indeed, to develop (and
sell!)  safety-critical systems, tools, such as Atelier B, need to pass a costly
certification process. Of course, this prevents continuous evolution of these
components. However, recent development in verification technologies, such as
other satisfiability modulo-theories (SMT)
solvers~\cite{DBLP:conf/cav/BarrettCDHJKRT11,DBLP:conf/tacas/MouraB08} has
resulted in significant progress. Therefore, we consider the time is ripe to
evaluate, through an open source project, the potential contribution of
incorporating such technologies in the tool set.  BEval is our contribution
towards this goal: an Atelier B plug-in that provides additional verification
engines and can be used for different utilities like: a disprover searching
counterexamples \cite{DBLP:journals/tsi/BendispostoLLS08}, a theorem prover
verifying the proof obligations and a mathematical rule validator tool checking
new reusable rules.

Besides, there are new requirements to the verification process in the B method.
We present our case study that produces proof obligations that cannot be
verified automatically with Atelier B built-in provers, and are very difficult
to show interactively, probably because these involve complex expressions with
bit vectors and math operators.

This paper is organized as follows. Section~\ref{sec:beval} presents BEval and
its context. Then section~\ref{sec:experiments} provides an experimental
evaluation of BEval based on a case study. We conclude with remarks in the last
section.

\section{The BEval Plug-in}
\label{sec:beval}

BEval\footnote{The BEval and its video demonstration are available at:
  https://github.com/ValerioMedeiros/BEval} is an open source tool to
systematize the verification of B expressions in Atelier B by integrating
ProB. BEval allows one to select B expressions and submits them for evaluation with
ProB. When the expression is true, BEval creates a matching proof rule
compatible with Atelier B built-in provers which may then discharge them
automatically.

\subsection{BEval's Architecture }

\begin{figure}[h!]
\centering \includegraphics[scale=0.37]{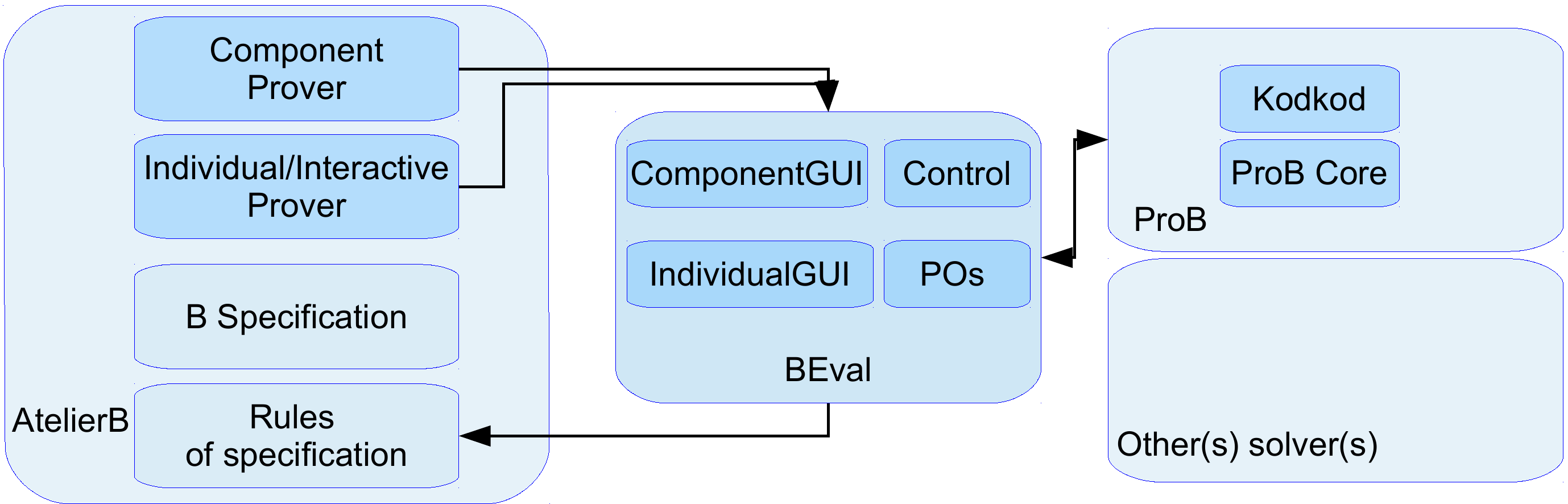}
\caption{An overview of component architecture.} 
\label{fig:architecture1}
\end{figure}

A B development generates a set of proof obligations. First, these proof
obligations are analyzed with Atelier B built-in automatic provers.  Then the
remaining proof obligations can be evaluated individually either by Atelier B
interactive prover or by BEval. The architecture of the verification framework
augmented with BEval is presented in Figure~\ref{fig:architecture1}. BEval
provides two graphical user interfaces: a rule evaluator that analyzes only one
proof obligation and a component evaluator that analyzes a set of selected proof
obligations.

Internally, BEval is composed of different components. They are:
\begin{itemize}

\item \verb'IndividualGUI' provides the graphical user interface for evaluation
  of an expression from within the interactive prover of Atelier B. This
  expression can be a full proof obligation, just a hypothesis of proof
  obligation, or a rule that embodies an important logic rule. Once verified,
  the tool offers to add the rule to the set of rules of the current
  specification in the interactive prover.

\item \verb'ComponentGUI' provides the graphical user interface for evaluation
  of a set of proof obligations.  This graphical interface contains different
  text areas: ProB evaluation parameters; results of evaluations; a list of
  proof obligations with its current state (proved/unproved).

\item \verb'Control' is responsible for controlling the communication between
  the tools. The communication between Atelier B, BEval and ProB is simple and
  uses command-line arguments from Java. Basically, Atelier B invokes a shell
  script passing as arguments information such as the path of module and the expression to
  evaluate. The shell script invokes BEval that redirects the output to a
  graphical user interface and calls a ProB client in the background.

\item \verb'POs' is responsible for managing the proof obligations stored in
  Atelier B format. This component is able to import one proof obligation or a
  set of proof obligations; and to export a set of true rules compatible with
  Atelier B.

\end{itemize}

\subsection{Graphical User Interface for Evaluating Proof Obligations} 

\begin{figure}[h] 
\centering
\includegraphics[scale=0.34]{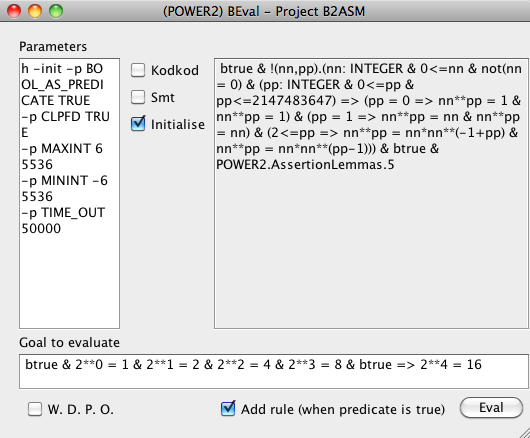} \ 
\includegraphics[scale=0.343]{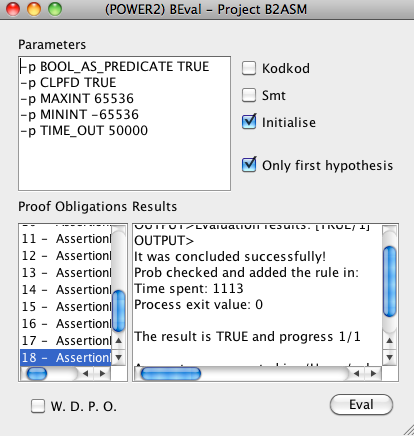} 
\caption{The graphical user interfaces of BEval:  on the left, the graphical interface
to submit one expression to evaluation; on the right, the graphical interface
to submit a set of proof obligations.}
\label{fig:gui}
\end{figure}

Figure~\ref{fig:gui} shows the two graphical user interfaces contributed by
BEval.  The left graphical interface can be invoked from the interactive prover
of Atelier B.  It has the following elements:
\begin{itemize}
\item \textit{Parameters} - located on the top-left of the window, it is an
  editable text where the user has access to the options used to call ProB;
\item \textit{Main options} - located on the top-middle of the window, three
  basic options are checkable; \textit{Kodkod} indicates that ProB may use the
  tool of the same name; \textit{Smt} indicates that ProB uses a more aggressive
  constraint solver; \textit{Initialise} indicates that definitions from the B
  component shall be loaded;
\item \textit{Hypothesis} - located on the top-right of the window, it presents
  the hypothesis that the user may want to add to the goal, addition of such
  hypothesis shall be performed with copy-and-paste operations; 
\item \textit{Goal to evaluate} - it is an editable text that contains the
  expression that will be sent to ProB; 
\item \textit{Add rule} - if this option is checked, whenever the goal evaluates
  to ``true'', a rule is generated and added into the corresponding \texttt{pmm}
  file \footnote{Atelier-B associates to each project a \texttt{pmm} file, where
    additional proof rules may be stored for use by the automatic provers to 
    discharge proof obligations.};
\item \textit{W.D.P.O.} - if that option is checked, then, whenever the goal
  evaluates to ``true'', the generated rule will be added to the corresponding
  \texttt{wd\_pmm} \footnote{The \texttt{wd\_pmm} file has a role similar to the
    \texttt{pmm} file but is used to discharge well-definedness proof
    obligations.}, otherwise it will be added to the common \verb'pmm' file.
\item \textit{Eval} - this button provokes the call to ProB on the current goal
  with the given list of parameters.
\end{itemize}
The right graphical interface is used within the components window. It is
similar, but has additional elements. First, \textit{Proof obligations} contains
a list of proof obligations and only those selected are evaluated. By default,
the selected items are unproved proof obligations. Second, \textit{Result} is a
text area that contains the output results of ProB's evaluations. The remaining
buttons are related to ProB parameters and are explained in the next section.

\subsection{Evaluation Parameters}

The options and parameters to the verification process are crucial and the
following are used by default in BEval:

\begin{itemize}

\item \textbf{-p MAXINT 65536 -p MININT -65536} sets the range for integers.

\item \textbf{-p init} loads definitions from B module.  This parameter is
  useful when the proof obligation was not fully expanded in only logic and math
  definitions. This parameter must be used when the proof obligation has a
  dependency of definitions.  For example, the proof obligation
  ``$[0,0,0,0,0,0,0,0]\in BYTE$" depends on the definition $BYTE=(1..8
  \rightarrow \{0,1\})$, so that this proof obligation is expanded to
  ``$[0,0,0,0,0,0,0,0] \in (1..8 \rightarrow \{0,1\})$'' and becomes
  independent. 

\item \textbf{-p KODKOD TRUE} indicates that ProB may use a constraint solver
  for relational logic, called Kodkod. This parameter allows a mixture of
  SAT-solving and ProB's own constraint-solving capabilities according to
  \cite{DBLP:conf/fm/PlaggeL12}.

\item \textbf{-p TIME\_OUT} sets the run time budget for the constraint solver.
 
\item \textbf{-p SMT TRUE} forces ProB to do more aggressive constraint solving.
 
\item \textbf{-p CLPFD TRUE} enables constraint logic programming over finite
  domains. It restricts range to $(-2^{28}..2^{28-1})$ on 32 bit computers.


\end{itemize}

Additional parameters are available, the full list being available in ProB's web
site~\cite{ProB_Manual}.

\subsection{Adding Rules}

A rule is a formula added as an axiom in the prover's theories by being stored
in a \texttt{pmm} file associated to a B component. The created rules can be
reused to solve several proof obligations. These rules can be added individually
by \verb'IndividualGUI' or several rules can be added by
\verb'ComponentGUI'. The following rule is very simple and it was generated by
BEval containing the information: name, date, spent time and the rule composed
by hypothesis and goal.

{\footnotesize
\begin{verbatim}
THEORY RulesProBAssertionLemmas_1 IS 
  /* Expression from (AssertionLemmas_1), it was added  in Thu Jun 27 18:02:32 BRT 2013
  evaluated with ProB in 5913 milliseconds. Module Path:/B_Resources/BYTE_DEFINITION.mch */	 
  "`Check assertion (card(BYTE) = 256) deduction - ref 3.2, 4.2, 5.3'"
  BYTE = (1..8 --> {0,1}) =>   (card(BYTE) = 256)
END
\end{verbatim}
}

Each created true rule has a relation with one proof obligation. When the
interactive prover of Atelier B is evaluating a proof obligation and
BEval-\verb'IndividualGUI' created a rule then the interactive prover can apply
the created rule in the evaluation of the current proof obligation.

BEval-\verb'ComponentGUI' creates a set of rules and a set of ``\textit{User
  Pass}", that is a sequence of proof commands. A \textit{User Pass} can be used
in automatic prover and can indicate a rule to apply in the selected proof
obligation. Each created \textit{User Pass} selects the proof obligations
by name and defines the rule to invoke. The following \textit{User Pass}
example selects the proof obligation named of \textit{initialisation} and
invokes the prover using the rule named of ``Rule1".

{\footnotesize
\begin{verbatim}
THEORY User_Pass IS
        Operation(Initialisation) & mp(Tac(RulesProBAssertionLemmas_1))
END
\end{verbatim}
}

\section{Experiments}
\label{sec:experiments}

We used Atelier-B to develop a reusable set of basic definitions to model
hardware concepts, data types concepts and a micro-controller instruction
set~\cite{Valerio_SBMF2012}.  These definitions are grouped into separated
development projects and are available as libraries.

The following table presents only the results of the most basic components using
the default parameters of BEval.  The components \textit{Power} and
\textit{Power2} contain the standard definition of exponentiation and it is
essential to establish the relationship between bit vectors and integer
arithmetics. The components \textit{BIT}, \textit{BYTE} and \textit{BV16} define
bit, bit vectors with size 8 and 16, basic functions to manipulate bit vectors
and important lemmas.

The proof obligations are classified in two groups: common and W.D.
(well-definedness proof obligations). The columns represent respectively:
\textbf{T. POs}, total number of proof obligations; \textbf{F1}, number of
verified proof obligations with force 1 of Atelier B's prover;
\textbf{F1;F2;F3}, number of verified proof obligations with force
3\footnote{Higher forces use mechanisms consuming more time, CPU and memory
resources.} after applying force 1 and 2; \textbf{F1;F2;F3;BEVAL}, number of
verified proof obligations with BEval after applying forces 1, 2 and 3;
\textbf{Gain}, percentage of proof obligations verified automatically and
exclusively by BEval. The symbol ``-'' represents no changes in the number of
verified proof obligations compared to the last applied strategy.

\begin{center}
  \begin{tabular}{| c | c | c | c | c | c | c |  c | c | c | c |}
  \hline
 &  \multicolumn{5}{|c|}{\textbf{Common POs}	} &   \multicolumn{5}{|c|}{\textbf{W. D. P. Os.}}   \\
  \hline						
\scriptsize{\textbf{Name}} & \scriptsize{\textbf{T. POs} }& \scriptsize{ \textbf{F1} }&\scriptsize{\textbf{F1;F2;F3}}&\scriptsize{\textbf{F1;F2;F3;BEval}} & \scriptsize{\textbf{Gain}} & \scriptsize{\textbf{T.	POs}} & \scriptsize{\textbf{F1}} &	\scriptsize{\textbf{F1;F2;F3}} & \scriptsize{\textbf{F1;F2;F3;BEval}} & \scriptsize{\textbf{Gain}}\\  \hline
Power & 3				&   2	&      -		&	   -	&		0\% & 	4	&  4 & 	- & 	-	          & 0\%  \\   \hline
Power2 & 18			&   2	&      -		&	   18	&		88\% & 	0	&  - & 	- & 	-	          & 0\%  \\   \hline
BIT & 49			&   23	&      -		&	   49	&		53\% & 	 69 	&  30 & 	- & 	-	          & 0\%  \\   \hline
BYTE & 18			&   12	&      -		&	   18	&		33\% & 	 136 	&  129 &  - & 	132	   & 2\% \\   \hline
BV16 & 6			&   2	&      -		&	   6	&		66\% & 	69 	&  67 &  - & 	69	   & 2\%  \\   \hline
  \end{tabular}
\end{center}

Almost all components have significant gains using BEval. This is significative
since it relieves the developer from the burden of manually verifying a
significant percentage of proof obligations and helps him focus on the more
interesting proof obligations and ultimately benefits his productivity. 

However, there are still some issues and limitations. Differences in the B
syntax supported by Atelier B and ProB need to be fixed to support the
evaluation of all components of micro-controller \cite{Valerio_SBMF2012}. 


\section{Conclusion and Perspectives} \label{sec:conclusions} 

Finally, BEval is a tool able to import proof obligations from Atelier B, and
convert and submit them for evaluation to ProB, interprets the results of that
evaluation and create proof rules in Atelier-B accordingly. BEval's integration
allows to exploit different strategies from the theorem prover of Atelier B and
constraint logic solver of ProB.  The results obtained with the verification of
hardware library demonstrates a better ability of the constraint logic solver of
ProB than the theorem prover of Atelier B for manipulating a class of
expressions. The results presented in this paper show again that providing a
port-folio of complementary provers is an effective approach to improve IDEs for
formal development.

Another related tool is the Rodin SMT Plug-in~\cite{DBLP:conf/asm/DeharbeFGV12},
this plug-in supports proof obligations generated from event-B specifications
and converts them to SMT format. In the future, BEval can also be integrated to
Rodin SMT Plug-in and exploit its abilities. Alternatively, the current SMT
translator of ProB~\cite{DBLP:conf/fm/PlaggeL12} can be improved and integrated
with news SMT solvers.

There are several possible new features and improvements for BEval. The small
differences in B parsers of Atelier B and ProB can be solved by creating a
pre-parser. Besides, ProB also has some limitations related to B constructs
supported by Atelier B, but these limitations are being solved; also other tools
may be investigated.

\paragraph{Acknowledgment.} The evaluation of the model would not have been
possible without the help of Michael Leuschel who kindly provided feedback and
developed improvements to ProB to meet our needs.

\bibliographystyle{eptcs}
\bibliography{paper}

\begin{thebibliography}{10}
\providecommand{\bibitemdeclare}[2]{}
\providecommand{\surnamestart}{}
\providecommand{\surnameend}{}
\providecommand{\urlprefix}{Available at }
\providecommand{\url}[1]{\texttt{#1}}
\providecommand{\href}[2]{\texttt{#2}}
\providecommand{\urlalt}[2]{\href{#1}{#2}}
\providecommand{\doi}[1]{doi:\urlalt{http://dx.doi.org/#1}{#1}}
\providecommand{\bibinfo}[2]{#2}

\bibitemdeclare{book}{DBLP:books/daglib/0015096}
\bibitem{DBLP:books/daglib/0015096}
\bibinfo{author}{Jean-Raymond \surnamestart Abrial\surnameend}
  (\bibinfo{year}{2005}): \emph{\bibinfo{title}{The B-book - assigning programs
  to meanings}}.
\newblock \bibinfo{publisher}{Cambridge University Press}.

\bibitemdeclare{article}{AbrialBHHMV10}
\bibitem{AbrialBHHMV10}
\bibinfo{author}{Jean-Raymond \surnamestart Abrial\surnameend},
  \bibinfo{author}{Michael \surnamestart Butler\surnameend},
  \bibinfo{author}{Stefan \surnamestart Hallerstede\surnameend},
  \bibinfo{author}{Thai~Son \surnamestart Hoang\surnameend},
  \bibinfo{author}{Farhad \surnamestart Mehta\surnameend} \&
  \bibinfo{author}{Laurent \surnamestart Voisin\surnameend}
  (\bibinfo{year}{2010}): \emph{\bibinfo{title}{Rodin: an open toolset for
  modelling and reasoning in {Event-B}}}.
\newblock {\sl \bibinfo{journal}{STTT}}
  \bibinfo{volume}{12}(\bibinfo{number}{6}), pp. \bibinfo{pages}{447--466}.
\newblock \urlprefix\url{http://dx.doi.org/10.1007/s10009-010-0145-y}.

\bibitemdeclare{inproceedings}{DBLP:conf/cav/BarrettCDHJKRT11}
\bibitem{DBLP:conf/cav/BarrettCDHJKRT11}
\bibinfo{author}{Clark \surnamestart Barrett\surnameend},
  \bibinfo{author}{Christopher~L. \surnamestart Conway\surnameend},
  \bibinfo{author}{Morgan \surnamestart Deters\surnameend},
  \bibinfo{author}{Liana \surnamestart Hadarean\surnameend},
  \bibinfo{author}{Dejan \surnamestart Jovanovic\surnameend},
  \bibinfo{author}{Tim \surnamestart King\surnameend}, \bibinfo{author}{Andrew
  \surnamestart Reynolds\surnameend} \& \bibinfo{author}{Cesare \surnamestart
  Tinelli\surnameend} (\bibinfo{year}{2011}): \emph{\bibinfo{title}{CVC4}}.
\newblock In: {\sl \bibinfo{booktitle}{CAV}}, pp. \bibinfo{pages}{171--177}.
\newblock \urlprefix\url{http://dx.doi.org/10.1007/978-3-642-22110-1_14}.

\bibitemdeclare{article}{DBLP:journals/tsi/BendispostoLLS08}
\bibitem{DBLP:journals/tsi/BendispostoLLS08}
\bibinfo{author}{Jens \surnamestart Bendisposto\surnameend},
  \bibinfo{author}{Michael \surnamestart Leuschel\surnameend},
  \bibinfo{author}{O.~\surnamestart Ligot\surnameend} \&
  \bibinfo{author}{Mireille \surnamestart Samia\surnameend}
  (\bibinfo{year}{2008}): \emph{\bibinfo{title}{La validation de mod{\`e}les
  Event-B avec le plug-in ProB pour RODIN}}.
\newblock {\sl \bibinfo{journal}{Technique et Science Informatiques}}
  \bibinfo{volume}{27}(\bibinfo{number}{8}), pp. \bibinfo{pages}{1065--1084}.
\newblock \urlprefix\url{http://dx.doi.org/10.3166/tsi.27.1065-1084}.

\bibitemdeclare{manual}{AtelierB}
\bibitem{AtelierB}
\bibinfo{organization}{ClearSy System Engineering},
  \bibinfo{address}{Aix-en-Provence}: \emph{\bibinfo{title}{Atelier B - User
  Manual}}.
\newblock
  \urlprefix\url{http://www.atelierb.eu/manuels/manuel-utilisateur-atelier-b-4%
.0-en.pdf}.

\bibitemdeclare{inproceedings}{DBLP:conf/asm/DeharbeFGV12}
\bibitem{DBLP:conf/asm/DeharbeFGV12}
\bibinfo{author}{David \surnamestart D{\'e}harbe\surnameend},
  \bibinfo{author}{Pascal \surnamestart Fontaine\surnameend},
  \bibinfo{author}{Yoann \surnamestart Guyot\surnameend} \&
  \bibinfo{author}{Laurent \surnamestart Voisin\surnameend}
  (\bibinfo{year}{2012}): \emph{\bibinfo{title}{SMT Solvers for Rodin}}.
\newblock In: {\sl \bibinfo{booktitle}{ABZ}}, pp. \bibinfo{pages}{194--207}.
\newblock \urlprefix\url{http://dx.doi.org/10.1007/978-3-642-30885-7_14}.

\bibitemdeclare{misc}{ProB_Manual}
\bibitem{ProB_Manual}
\bibinfo{author}{Michael \surnamestart Leuschel\surnameend}
  (\bibinfo{year}{2011}): \emph{\bibinfo{title}{User Manual}}.
\newblock
  \urlprefix\url{http://www.stups.uni-duesseldorf.de/ProB/index.php5/User\_Man%
ual}.

\bibitemdeclare{inproceedings}{DBLP:dblp_conf/fm/LeuschelB03}
\bibitem{DBLP:dblp_conf/fm/LeuschelB03}
\bibinfo{author}{Michael \surnamestart Leuschel\surnameend} \&
  \bibinfo{author}{Michael~J. \surnamestart Butler\surnameend}
  (\bibinfo{year}{2003}): \emph{\bibinfo{title}{ProB: A Model Checker for B.}}
\newblock In: {\sl \bibinfo{booktitle}{FME}}, pp. \bibinfo{pages}{855--874}.
\newblock \urlprefix\url{http://dx.doi.org/10.1007/978-3-540-45236-2_46}.

\bibitemdeclare{article}{LiBeLe07_219}
\bibitem{LiBeLe07_219}
\bibinfo{author}{Olivier \surnamestart Ligot\surnameend}, \bibinfo{author}{Jens
  \surnamestart Bendisposto\surnameend} \& \bibinfo{author}{Michael
  \surnamestart Leuschel\surnameend} (\bibinfo{year}{2007}):
  \emph{\bibinfo{title}{Debugging Event-B Models using the ProB Disprover
  Plug-in}}.
\newblock {\sl \bibinfo{journal}{Proceedings AFADL'07}}.

\bibitemdeclare{inproceedings}{Valerio_SBMF2012}
\bibitem{Valerio_SBMF2012}
\bibinfo{author}{Val\'{e}rio~G. \surnamestart Medeiros{\ }Jr.\surnameend} \&
  \bibinfo{author}{David \surnamestart D\'{e}harbe\surnameend}
  (\bibinfo{year}{2012}): \emph{\bibinfo{title}{{E}xperience in {M}odeling a
  {M}icrocontroller {I}nstruction {S}et {U}sing {B}}}.
\newblock In: {\sl \bibinfo{booktitle}{Brazilian Symposium on Formal Methods}},
  \bibinfo{organization}{SBMF}, \bibinfo{address}{Natal - RN}.

\bibitemdeclare{inproceedings}{DBLP:conf/tacas/MouraB08}
\bibitem{DBLP:conf/tacas/MouraB08}
\bibinfo{author}{Leonardo~Mendon\c{c}a \surnamestart de~Moura\surnameend} \&
  \bibinfo{author}{Nikolaj \surnamestart Bj{\o}rner\surnameend}
  (\bibinfo{year}{2008}): \emph{\bibinfo{title}{Z3: An Efficient SMT Solver}}.
\newblock In: {\sl \bibinfo{booktitle}{TACAS}}, pp. \bibinfo{pages}{337--340}.
\newblock \urlprefix\url{http://dx.doi.org/10.1007/978-3-540-78800-3_24}.

\bibitemdeclare{inproceedings}{DBLP:conf/fm/PlaggeL12}
\bibitem{DBLP:conf/fm/PlaggeL12}
\bibinfo{author}{Daniel \surnamestart Plagge\surnameend} \&
  \bibinfo{author}{Michael \surnamestart Leuschel\surnameend}
  (\bibinfo{year}{2012}): \emph{\bibinfo{title}{Validating B, Z and TLA + Using
  ProB and Kodkod}}.
\newblock In: {\sl \bibinfo{booktitle}{FM}}, pp. \bibinfo{pages}{372--386}.
\newblock \urlprefix\url{http://dx.doi.org/10.1007/978-3-642-32759-9_31}.

\end{thebibliography}
\end{document}